\documentclass[preprintnumbers,amsmath,amssymbm,prd]{revtex4}
\usepackage{epsfig}
\usepackage{graphicx}

\begin{document}
\title{The Reissner-Nordstr\"om black hole with the fastest relaxation rate}
\author{Shahar Hod}
\affiliation{The Ruppin Academic Center, Emeq Hefer 40250, Israel}
\affiliation{ }
\affiliation{The Hadassah Institute, Jerusalem 91010, Israel}
\date{\today}

\begin{abstract}
\ \ \ Numerous {\it numerical} investigations of the quasinormal
resonant spectra of Kerr-Newman black holes have revealed the
interesting fact that the characteristic relaxation times
$\tau({\bar a},{\bar Q})$ of these canonical black-hole spacetimes
can be described by a two-dimensional function ${\bar \tau}\equiv
\tau/M$ which increases monotonically with increasing values of the
dimensionless angular-momentum parameter ${\bar a}\equiv J/M^2$ and,
in addition, is characterized by a non-trivial ({\it non}-monotonic)
functional dependence on the dimensionless charge parameter ${\bar
Q}\equiv Q/M$. In particular, previous numerical investigations have
indicated that, within the family of spherically symmetric charged
Reissner-Nordstr\"om spacetimes, the black hole with ${\bar Q}\simeq
0.7$ has the {\it fastest} relaxation rate. In the present paper we
use {\it analytical} techniques in order to investigate this
intriguing non-monotonic functional dependence of the
Reissner-Nordstr\"om black-hole relaxation rates on the
dimensionless physical parameter ${\bar Q}$. In particular, it is
proved that, in the eikonal (geometric-optics) regime, the black
hole with ${\bar Q}={{\sqrt{51-3\sqrt{33}}}\over{8}}\simeq 0.73$ is
characterized by the {\it fastest} relaxation rate (the smallest
dimensionless relaxation time ${\bar \tau}$) within the family of
charged Reissner-Nordstr\"om black-hole spacetimes.
\end{abstract}
\bigskip
\maketitle

\section{Introduction}

The mathematically elegant uniqueness theorems of Israel, Carter,
and Hawking \cite{Isr,Car,Haw1} (see also \cite{Rob,Isp}) have
revealed the physically important fact that all asymptotically flat
stationary black-hole solutions of the non-linearly coupled
Einstein-Maxwell field equations belong to the Kerr-Newman family of
curved spacetimes \cite{Chan,Kerr,Newman,Notehh,Hodrc,Herkr}. The
three-dimensional phase-space of these canonical black-hole
spacetimes is described by conserved physical parameters that can be
measured by asymptotic observers: the black-hole mass $M$, the
black-hole electric charge $Q$, and the black-hole angular momentum
$J\equiv Ma$.

In accord with the results of the uniqueness theorems
\cite{Isr,Car,Haw1,Rob,Isp}, the linearized dynamics of massless
gravitational and electromagnetic fields in perturbed Kerr-Newman
black-hole spacetimes are characterized by damped quasinormal
resonant modes with the physically motivated boundary conditions of
purely ingoing waves at the black-hole outer horizon and purely
outgoing waves at spatial infinity \cite{Det}. These exponentially
decaying black-hole-field oscillation modes describe the gradual
dissipation of the external massless perturbation fields which are
radiated into the black hole and to spatial infinity. As a result of
this dissipation process, the perturbed spacetime gradually returns
into a stationary Kerr-Newman black-hole solution of the
Einstein-Maxwell field equations.

Interestingly, for a given set $\{{\bar Q}\equiv Q/M, {\bar a}\equiv
J/M^2\}$ of the Kerr-Newman dimensionless physical parameters
\cite{Noteunit,Noteqq}, the physically motivated boundary conditions
at the black-hole horizon and at spatial infinity single out a
discrete set $\{{\bar \omega}\equiv M\omega({\bar Q},{\bar
a};l,m,n)\}^{n=\infty}_{n=0}$ \cite{Noteml} of dimensionless complex
resonant frequencies which characterize the relaxation dynamics of
the massless fields in the curved black-hole spacetime (see
\cite{Nol,Ber,Kon} for excellent reviews on the physically exciting
topic of black-hole resonant spectra). In particular, the
fundamental resonant frequency ${\bar \omega(n=0)}$ \cite{Notefun}
determines, through the relation
\begin{equation}\label{Eq1}
{\bar \tau_{\text{relax}}} \equiv {\bar\omega_{\text{I}}}^{-1}(n=0)\ ,
\end{equation}
the characteristic dimensionless relaxation time of the perturbed
black-hole spacetime.

One can ask: Within the family of charged and rotating Kerr-Newman
black-hole spacetimes, which black hole has the longest relaxation
time ${\bar \tau_{\text{relax}}}({\bar Q},{\bar a})$? Interestingly,
it has previously been proved that the answer to this question is
{\it not} unique. In particular, previous analytical
\cite{Ka1,Ka2,Ka3,Ka4} and numerical \cite{KN1,KN2,KN3,KN4,KN5}
studies of the Kerr and Kerr-Newman quasinormal resonance spectra
have explicitly proved that the characteristic black-hole relaxation
times grow unboundedly as the extremal limit ${\bar T_{\text{BH}}}\to 0$ is approached
\cite{Notealnm}:
\begin{equation}\label{Eq2}
{\bar \tau_{\text{relax}}}\propto{\bar
T_{\text{BH}}}^{-1}\to\infty\ \ \ \ \text{for}\ \ \ \ {\bar
T_{\text{BH}}}\to 0\  ,
\end{equation}
where ${\bar T_{\text{BH}}}\equiv MT_{\text{BH}}$ is the
characteristic dimensionless Bekenstein-Hawking temperature of the
black-hole spacetime \cite{Notetep,Bekt,Hawt}.

In the present paper we shall focus on the opposite regime of
Reissner-Nordstr\"om black holes which {\it minimize} the
dimensionless relaxation time (\ref{Eq1}). In particular, we here
raise the following physically interesting question: Within the
family of spherically symmetric charged Reissner-Nordstr\"om black
holes, which black hole has the {\it fastest} relaxation rate [that
is, the shortest relaxation time ${\bar \tau_{\text{relax}}}({\bar
Q})$]? Intriguingly, below we shall explicitly prove that, as
opposed to the case of near-extremal spacetimes which maximize the
black-hole relaxation times [see Eq. (\ref{Eq2})], the answer to the
black-hole minimization question is {\it unique}.

\section{Review of former analytical and numerical studies}

Former analytical and numerical studies of the Kerr-Newman resonance
spectra \cite{Ka1,Ka2,Ka3,Ka4,KN1,KN2,KN3,KN4,KN5} have revealed the
fact that, for a given value of the black-hole electric charge
${\bar Q}$, the characteristic black-hole relaxation time ${\bar
\tau_{\text{relax}}}({\bar Q},{\bar a})$ is a monotonically
increasing function of the rotation parameter ${\bar a}$. Hence,
within the canonical family of Kerr-Newman black-hole spacetimes,
the relaxation rate ${\bar \tau_{\text{relax}}}^{-1}({\bar Q},{\bar
a})$ can be maximized by considering non-spinning (${\bar a}=0$)
black holes \cite{Noteqdf}. In the present paper we shall therefore
focus our attention on spherically symmetric charged
Reissner-Nordstr\"om black-hole spacetimes.

In addition, former numerical studies \cite{KN1,KN4,KN5,RN1,RN2,RN3} of
the resonance spectra of charged black holes have revealed the interesting fact
that the black-hole relaxation times ${\bar
\tau_{\text{relax}}}({\bar Q},{\bar a})$ are characterized by a
non-trivial ({\it non}-monotonic) functional dependence on the
dimensionless physical parameter ${\bar Q}$. In particular, detailed
numerical computations \cite{KN2,RN3} indicate that, in the
physically interesting case of coupled gravitational-electromagnetic
quadrupole perturbations fields with $l=2$, the black-hole
relaxation time (\ref{Eq1}) is minimized for
\begin{equation}\label{Eq3}
{\bar Q}_{\text{min}}\simeq 0.7\ \ \ \ \text{for}\ \ \ \ l=2\  .
\end{equation}

The main goal of the present compact paper is to shed light on the
intriguing non-monotonic functional dependence of the characteristic
Reissner-Nordstr\"om black-hole relaxation times ${\bar
\tau_{\text{relax}}}({\bar Q})$ on the dimensionless charge
parameter ${\bar Q}$. In particular, in order to facilitate a fully
{\it analytical} treatment of the black-hole minimization problem,
we shall study in the present paper the eikonal (geometric-optics)
regime of the quasinormal resonance spectra which characterize the
composed black-hole-field system. Interestingly, below we shall
explicitly demonstrate that the analytical results obtained in the
eikonal $l\gg1$ regime agree remarkably well with the numerical
results \cite{KN2,RN3} of the quadrupole $l=2$ perturbation fields.

\section{The composed black-hole-field quasinormal resonance spectra in the eikonal (geometric-optics) $l\gg1$ regime}

We consider a Reissner-Nordstr\"om black-hole spacetime of mass $M$
and electric charge $Q$ which is characterized by the curved line
element \cite{Chan,Noteunit}
\begin{equation}\label{Eq4}
ds^2=-f(r)dt^2+{1\over{f(r)}}dr^2+r^2(d\theta^2+\sin^2\theta
d\phi^2)\  ,
\end{equation}
where
\begin{equation}\label{Eq5}
f(r)=1-{{2M}\over{r}}+{{Q^2}\over{r^2}}\  .
\end{equation}
The radii \cite{Chan}
\begin{equation}\label{Eq6}
r_{\pm}=M\pm\sqrt{M^2-Q^2}\
\end{equation}
of the inner and outer horizons of the charged black-hole spacetime
(\ref{Eq4}) are determined by the zeroes of the radially-dependent
metric function (\ref{Eq5})

Massless perturbations fields of the black-hole spacetime (\ref{Eq4})
are governed by the Schr\"odinger-like ordinary differential
equation \cite{Chan}
\begin{equation}\label{Eq7}
{{d^2\psi}\over{dy^2}}+V_l[r(y)]\psi=0\  ,
\end{equation}
where the tortoise radial coordinate $y$ is defined by the
differential relation
\begin{equation}\label{Eq8}
dy={{dr}\over{f(r)}}\
\end{equation}
and, in the eikonal (geometric-optics) $l\gg1$ limit, the effective
black-hole-field curvature potential in (\ref{Eq7}) is given by the functional relation \cite{Chan}
\begin{equation}\label{Eq9}
V_l(r)=\omega^2-l(l+1){{f(r)}\over{r^2}}\  .
\end{equation}

The quasinormal resonant modes (damped oscillations) which
characterize the linearized relaxation dynamics of the perturbed
black-hole spacetime are determined by the Schr\"odinger-like radial
equation (\ref{Eq7}) with the physically motivated boundary conditions of purely
ingoing waves at the black-hole event horizon [which is
characterized by $y\to -\infty$] and purely outgoing waves at
spatial infinity [which is characterized by $y\to \infty$]:
\begin{equation}\label{Eq10}
\psi \sim
\begin{cases}
e^{-i\omega y} & \text{\ for\ \ \ } r\rightarrow r_+\ \
(y\rightarrow -\infty)\ ; \\ e^{i\omega y} & \text{ for\ \ \ }
r\rightarrow \infty\ \ \ (y\rightarrow \infty)\  .
\end{cases}
\end{equation}

We shall now use analytical techniques in order to determine the
characteristic quasinormal resonant frequencies which characterize
the black-hole spacetime (\ref{Eq4}) in the eikonal large-$l$
regime. To this end, we shall use the results of the elegant WKB
analysis presented in \cite{WKB1,WKB2}, according to which the
quasinormal resonance spectrum which characterizes the
Schr\"odinger-like ordinary differential equation (\ref{Eq7}) with
the boundary conditions (\ref{Eq10}) is determined, in the eikonal
$l\gg1$ regime, by the simple leading-order resonance equation
\cite{WKB1,WKB2}
\begin{equation}\label{Eq11}
i{{V_0}\over{\sqrt{2V^{(2)}_0}}}=n+{1\over 2}+O(l^{-1})\ \ \ \ ; \ \
\ \ n=0,1,2,...
\end{equation}
Here the composed black-hole-field curvature potential $V_0$ and its
radial derivatives $V^{(k)}_0\equiv d^{k}V/dy^{k}$ [see Eq.
(\ref{Eq9})] are evaluated at the radial peak $y=y_0$, which is
characterized by the simple functional relation
\begin{equation}\label{Eq12}
{{dV}\over{dy}}=0\ \ \ \ \text{for}\ \ \ \ y=y_0\ \ (r=r_0)\  .
\end{equation}

Substituting the effective black-hole-field radial potential (\ref{Eq9}) into
(\ref{Eq12}), one finds the simple expression \cite{RN3}
\begin{equation}\label{Eq13}
{\bar r_0}\equiv{{r_0}\over{M}}={1\over2}\Big(3+\sqrt{9-8{\bar
Q}^2}\Big)\
\end{equation}
for the dimensionless radius which characterizes the peak of the
curvature potential in the eikonal large-$l$ regime. Substituting
(\ref{Eq13}) into the WKB resonance equation (\ref{Eq11}) and using
the differential relation (\ref{Eq8}), one obtains the (rather
cumbersome) functional expressions
\begin{equation}\label{Eq14}
{\bar \omega_{\text{R}}}={{(2l+1)\sqrt{2}\sqrt{3+\sqrt{9-8{\bar Q}^2}-2{\bar Q}^2}}
\over{\big[3+\sqrt{9-8{\bar Q}^2}\big]^2}}
\end{equation}
and
\begin{equation}\label{Eq15}
{\bar \omega_{\text{I}}}=-\Big(n+{1\over2}\Big)\cdot{{4\sqrt{3}
\sqrt{\Big[3+\sqrt{9-8{\bar Q}^2}-2{\bar Q}^2\Big]\Big[3+\sqrt{9-8{\bar Q}^2}-{8\over3}{\bar Q}^2\Big]}}
\over
{\big[3+\sqrt{9-8{\bar Q}^2}\big]^3}}\ \ \ \ ; \ \ \ \ n=0,1,2,...
\end{equation}
for the characteristic dimensionless resonant frequencies of the charged Reissner-Nordstr\"om
black-hole spacetime (\ref{Eq4}).

Inspecting the analytically derived geometric-optics expression
(\ref{Eq15}), one reveals the interesting fact that the function
${\bar\omega_{\text{I}}}({\bar Q})$ has a non-trivial
(non-monotonic) dependence on the dimensionless physical parameter
${\bar Q}$. In particular, analyzing the function
${\bar\omega^{(0)}_{\text{I}}}({\bar Q})$ [see Eq. (\ref{Eq15})] for
the fundamental ($n=0$) resonant frequency in the eikonal regime, we
find that the characteristic dimensionless relaxation time
${\bar\tau({\bar Q})}\equiv 1/{\bar\omega^{(0)}_{\text{I}}}$ of the
composed black-hole-field system is {\it minimized} for the
particular value
\begin{equation}\label{Eq16}
{\bar Q}_{\text{min}}={{\sqrt{51-3\sqrt{33}}}\over{8}}\simeq 0.726\
\ \ \ \text{for}\ \ \ \ l\gg1\
\end{equation}
of the dimensionless black-hole charge parameter.

Interestingly, one finds that the analytically derived expression
(\ref{Eq16}) for the dimensionless charge parameter ${\bar
Q}_{\text{min}}$, which characterizes the Reissner-Nordstr\"om black
hole with the fastest relaxation rate (the shortest relaxation time)
in the eikonal large-$l$ regime, agrees remarkably well with the
numerically computed value ${\bar
Q}^{\text{numerical}}_{\text{min}}\simeq0.7$ \cite{KN3,RN3} [see Eq.
(\ref{Eq3})] for the quadrupole $l=2$ perturbation fields.

\section{Summary and Discussion}

The characteristic quasinormal resonance spectra of black holes have
attracted the attention of physicists and mathematicians during the
last five decades \cite{Nol,Ber,Kon}. In the present paper we have
raised the following physically interesting questions: Within the
canonical family of charged and rotating Kerr-Newman black holes,
which black hole has the slowest relaxation rate and which black
hole has the {\it fastest} relaxation rate?

Using analytical techniques we have shown that, while the answer to
the first question is not unique [see the characteristic relation
(\ref{Eq2}) for the family of near-extremal black holes with
diverging relaxation times in the ${\bar T_{\text{BH}}}\to0$ limit],
the answer to the second question is unique within the phase space
of spherically symmetric charged Reissner-Nordstr\"om black-hole
spacetimes. In particular, based on previous analytical and
numerical studies of the black-hole resonance spectra
\cite{Ka1,Ka2,Ka3,Ka4,KN1,KN2,KN3,KN4,KN5} which revealed the
interesting fact that, for a given value of the dimensionless charge
parameter ${\bar Q}$, the characteristic black-hole relaxation time
${\bar \tau}({\bar a},{\bar Q})$ is a monotonically increasing
function of the dimensionless angular-momentum parameter ${\bar a}$,
we have focused our attention on the relaxation properties of
non-spinning (${\bar a}=0$) charged Reissner-Nordstr\"om black-hole
spacetimes. In addition, in order to facilitate a fully {\it
analytical} exploration of the intriguing non-monotonic functional
dependence of the black-hole relaxation rates ${\bar
\tau}^{-1}({\bar Q})$ on the dimensionless physical parameter ${\bar
Q}$, we have considered in the present paper the eikonal
(geometric-optics) regime of the composed black-hole-field resonance
spectrum.

Using standard analytical techniques \cite{WKB1,WKB2}, we have
explicitly demonstrated that the composed black-hole-field
relaxation spectrum is characterized by a non-trivial
(non-monotonic) functional dependence on the dimensionless charge
parameter ${\bar Q}$. This analytically derived result supports the
numerical data presented in \cite{KN2,RN3}. Moreover, it is
interesting to note that the {\it analytically} calculated
black-hole charge parameter [see Eq. (\ref{Eq16})]
\begin{equation}\label{Eq17}
\Big({{Q}\over{M}}\Big)^{\text{analytical}}_{\text{min}}={{\sqrt{51-3\sqrt{33}}}\over{8}}\
,
\end{equation}
which minimizes the dimensionless relaxation time of the composed
Reissner-Nordstr\"om-black-hole-field system in the eikonal $l\gg1$
regime, agrees remarkably well with the {\it numerically} computed
charge parameter ${\bar Q}^{\text{numerical}}_{\text{min}}\simeq0.7$
[see Eq. (\ref{Eq3})] which minimizes the black-hole relaxation time
in the canonical case of coupled gravitational-electromagnetic
quadrupole perturbation fields with $l=2$ \cite{KN2,RN3}.

\bigskip
\noindent
{\bf ACKNOWLEDGMENTS}
\bigskip

This research is supported by the Carmel Science Foundation. I would
like to thank Yael Oren, Arbel M. Ongo, Ayelet B. Lata, and Alona B.
Tea for helpful discussions.



\begin{thebibliography}{99}
\bibitem{Notehh} See \cite{Hodrc,Herkr} for the physically interesting case
of non Kerr-Newman black-hole solutions of the coupled
Einstein-Maxwell-scalar field equations which describe spinning
black holes that support linear \cite{Hodrc} and non-linear
\cite{Herkr} stationary scalar (bosonic) fields.
\bibitem{Noteunit} We shall use natural units in which $G=c=\hbar=1$.
\bibitem{Noteqq} We shall assume without loss of generality that ${\bar Q}\geq0$.
\bibitem{Noteml} The dimensionless parameters $\{l,m\}$ are respectively the spheroidal
harmonic index and the azimuthal harmonic index of the massless
perturbation modes.
\bibitem{Notefun} The fundamental resonant mode of the black-hole spacetime
is defined as the mode with the smallest imaginary part (the
perturbation mode with the largest relaxation time).
\bibitem{Notetep} The Bekenstein-Hawking temperature of the
Kerr-Newman black hole is given by the functional expression
$T_{\text{BH}}=(r_+-r_-)/4\pi(r^2_++a^2)$, where $r_{\pm}$ are the
horizon radii of the black-hole spacetime [see Eq. (\ref{Eq6})
below].
\bibitem{Noteqdf} It is worth noting that, for Kerr-Newman black holes,
the qualitative difference in the functional behavior of the
relaxation time between the case in which the charge is varied and
the case in which the spin parameter is varied is probably related
to the fact that the null circular geodesic of a near-extremal Kerr
black hole with $a/M\to 1^-$ is located in the vicinity of the
black-hole horizon, whereas the null circular geodesic of a
near-extremal Reissner-Nordstr\"om black hole with $Q/M\to 1^-$ is
located a finite radial distance from the black-hole horizon.
\end{thebibliography}
\end{document}